\documentclass[twocolumn,english,prl,showpacs]{revtex4}
\usepackage[T1]{fontenc}
\usepackage[latin1]{inputenc}
\usepackage{graphicx}
\usepackage{amsmath}
\usepackage{amsfonts}
\usepackage{amssymb}
\usepackage{mathrsfs}
\makeatletter

\usepackage{babel}
\makeatother

\begin{document}
\title{BCS-BEC Crossover in the Strongly Correlated Regime of
ultra-cold Fermi gases} \author{S. G. Bhongale}
\email{bhongale@phys.unm.edu} \affiliation{Department of Physics and
Astronomy, University of New Mexico, Albuquerque, NM 87131, USA.}
\author{S. J. J. M. F. Kokkelmans} \affiliation{Eindhoven University
of Technology, P.O. Box 513, 5600 Eindhoven, The Netherlands.}
\begin{abstract}
We study BCS-BEC crossover in the strongly correlated regime of two
component rotating Fermi gases. We predict that the strong
correlations induced by rotation will have the effect of modifying the
crossover region relative to the non-rotating situation. We show via
the two particle correlation function that the crossover smoothly
connects the $s$-wave paired fermionic fractional quantum Hall state
to the bosonic Laughlin state.
\end{abstract}
\date{\today}
\maketitle 

In recent years techniques based on Feshbach scattering resonances
\cite{fesh1,fesh2} in ultra-cold atomic gases have allowed the study
of condensation in a Fermi system
\cite{fermicond1,fermicond2,fermicond3}. For condensation to occur,
one can distinguish two distinct physical mechanisms: (1) formation of
bound pairs of fermionic atoms (molecules) which are composite bosons
and hence undergo Bose-Einstein condensation (BEC), and (2)
condensation of Bardeen-Cooper-Schrieffer (BCS) pairs in analogy with
low temperature superconductivity. In separate publications
\cite{eagles,leggett} both Eagles and Leggett argued that these
scenarios were limiting cases of a more general theory, the so-called
BCS-BEC crossover. It was only recently that this crossover phenomenon
was observed in rotating trap experiments via the use of a Feshbach
scattering resonance. A vortex lattice generated in the molecular BEC
phase was observed to persist into the BCS paired phase as the
interaction is adiabatically tuned from repulsive to attractive across
the Feshbach resonance \cite{crossvert}.

\begin{figure}[b]
  \includegraphics[scale=.38]{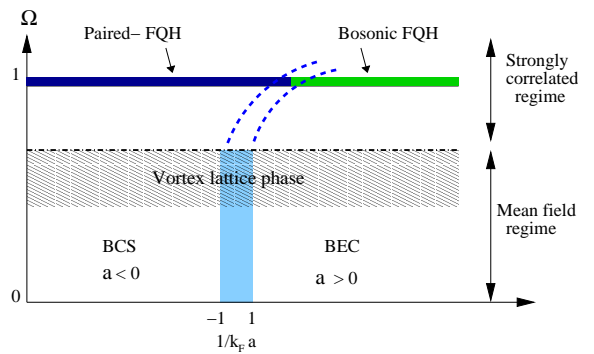}
\caption{Schematic of the zero temperature phase diagram of the two
component Fermi gas with Feshbach tuning parameter $1/k_Fa$ along the
horizontal axis and rotational frequency $\Omega$ (in units of
trapping frequency $\omega$) along the vertical axis. $a$ represents
the $s$-wave scattering length.}
\label{phasedig}
\end{figure}

While developments such as above have allowed us to enhance our
understanding of numerous many body effects, applications of trapped
atomic systems to study strong correlation effects such as those
responsible for the fractional quantum Hall (FQH) effect are
limited. However, there have been theoretical proposals to configure
the ultra-cold atomic system in the FQH regime
\cite{wilkins,zoller}. These proposals are based on rotating the trap
at frequencies close to the trapping frequency, $\omega$. This allows
us to draw a phase diagram of experimental parameters (rotational
frequency $\Omega$ and the Feshbach tuning parameter) and identify
separate regions corresponding BEC, BCS and FQH as shown in
Fig.~\ref{phasedig}. The region below the dot-dashed line can be quite
successfully described at the mean field level and hence we will refer
to it as the mean field regime. The vertical shaded area in the mean
field regime represents the BCS-BEC crossover where $|k_Fa|>1$ ($k_F$
is the Fermi wave vector and $a$ is the two body scattering
length). While the crossover has been experimentally explored only in
the mean field regime, on general grounds one would expect it to exist
even in the strongly correlated regime. However, in this regime the
correlations will induce modifications to the mean field crossover
physics discussed above. This is in part due to the global vortex
lattice structure imprinted on the system due to rotations, which
gives rise to an emergent length scale corresponding to the vortex
radius. While this effect may not be significant in the
mean-field-vortex-lattice (hashed) region of the phase diagram, at
high rotations, where the number of particles is comparable to the
number of vortices, implications to the crossover physics may be
drastic. At the same time, very recently FQH states have attracted
special attention due to their possible use in topological schemes of
quantum computation. While manipulating interaction has remained a
major difficulty, possibility of crossover in the FQH regime of atomic
ensembles may turn out to be of significant importance in such
schemes.

Therefore, the goal of this letter is to investigate the implications
of strong correlations due to rotations on the BCS-BEC crossover. We
also argue that the crossover is expected to be smooth for $s$-wave
interactions. We will verify this by considering a specific Cooper
paired FQH state on the BCS side. Finally we establish the nature of
the strongly correlated state on the BEC side of the crossover by a
numerical study of the two-particle correlation functions.

We consider a two component Fermi system consisting of a 50\%-50\%
mixture of fermionic atoms in different hyperfine states represented
by $|\uparrow\rangle$ and $|\downarrow\rangle$ confined by a 2D
rotating harmonic trap. The Hamiltonian for this system in the FQH
regime ($\Omega-\omega\rightarrow 0^-$) in the absence of interactions
is given by
\begin{equation}
H=\sum_{\sigma}\int\phi_{\sigma}^{\dagger}({\bf
r})\frac{1}{2m}\left[{\bf p}+{\bf A({\bf
r})}\right]^2\phi_{\sigma}({\bf r}) d{\bf r},
\end{equation} 
with $m$ the mass and ${\bf A}=m\omega y\hat{\text{x}}-m\omega
x\hat{\text{y}}$ is analogous to the vector potential associated with
the magnetic field in the electronic FQH effect, and $\phi_{\sigma}$
is the annihilation operator representing fermionic atoms with spin
$\sigma$. Now in order to simplify the above Hamiltonian we choose to
work in a frame where the vector potential ${\bf A}$ is gauged
out. This is done by performing the Chern-Simons transformation
\cite{zhang} by attaching gauge field $ a_\alpha ({\bf
r})=-\nu\hbar\sum_{\sigma}\int d^2{\bf
r}'\epsilon_{\alpha\beta}\rho_{\sigma}({\bf r})({\bf r}-{\bf
r}')_{\beta}/|{\bf r}-{\bf r}'|^2$ to each bare particle resulting in
\begin{equation}
H=\int\varphi_{\sigma}^{\dagger}({\bf r})\frac{1}{2m}\left[{\bf
p}+{\bf A({\bf r})}+{\bf a({\bf
r})}\right]^2\varphi_{\sigma}({\bf r}) d{\bf r},
\end{equation}
where $\varphi_{\sigma}$ is the annihilation operator and
$\rho_{\sigma}$ is the density of composite fermions of spin $\sigma$
and $1/\nu$ is the fraction of the FQH effect. The transformation is
such that the average gauge field $\bar{\bf a}$ \cite{lucjan} cancels
the external field, thus ${\bf A}({\bf r})+{\bf a}({\bf r})={\bf
A}({\bf r})+\bar{\bf a}({\bf r})+\delta{\bf a}({\bf r})=\delta{\bf
a}({\bf r})$. We can now write the interaction part of the above
Hamiltonian as $H_{\text{int}}=(1/2m)\sum_{\sigma}\int d^2{\bf
r}\varphi_{\sigma}^{\dagger}({\bf r})\left[2 {\bf p}\delta {\bf
a}({\bf r})+\delta{\bf a}({\bf r})^2\right]\varphi_{\sigma}({\bf r})
=H_1+H_2$, where
\begin{eqnarray}
H_1&=&-\frac{\nu\hbar}{m}\sum_{\sigma\sigma'}\int\int d^2{\bf
r}d^2{\bf r}'\varphi_{\sigma}^{\dagger}({\bf
r})p_{\alpha}\varphi_{\sigma}({\bf
r})\epsilon_{\alpha\beta}\nonumber\\ &&\times\frac{({\bf r}-{\bf
r}')_{\beta}}{|{\bf r}-{\bf r}'|^2}\delta\rho_{\sigma'}({\bf
r}'),\label{h1}\\
H_2&=&\frac{\nu^2\hbar^2}{2m}\sum_{\sigma\sigma'\sigma''}\int\int\int
d^2 {\bf r}d^2 {\bf r}'d^2 {\bf r}''\rho_\sigma({\bf r})\nonumber\\
&&\times\frac{({\bf r}-{\bf r}')}{|{\bf r}-{\bf r}'|^2}\frac{({\bf
r}-{\bf r}'')}{|{\bf r}-{\bf r}''|^2} \delta\rho_{\sigma'}({\bf
r}')\delta\rho_{\sigma''}({\bf r}''),\label{h2}
\end{eqnarray}
and $\rho_\sigma(\bf r)=\varphi_{\sigma}^{\dagger}({\bf r})
\varphi_{\sigma}({\bf r})$ and $\delta\rho_{\sigma}({\bf
r})=\rho_{\sigma}({\bf r})-\bar{\rho}$, with $\bar{\rho}$ the average
density. Thus we see that even in the absence of interactions between
the bare particles, the Chern-Simons transformation gives rise to a
two-body (contained in $H_1$ and $H_2$) and three-body contribution
(only contained in $H_2$). However, $H_1$ represents a coupling of the
density to the current $j_{\alpha}({\bf
r})=\varphi_{\sigma}^{\dagger}({\bf r})p_{\alpha}\varphi_{\sigma}({\bf
r})$ which is beleived to be important only near the boundary of the
sample. It can be shown that the two-body part of $H_2$ is logarithmic
and has been attributed to have important consequences for the
formation of pairing in the electronic FQH effect \cite{lucjan}. The
three-body part will be neglected in the following, while the two-body
part can be taken as an additional contribution to the two-particle
atomic interaction.

\begin{figure}
  \includegraphics[scale=.30]{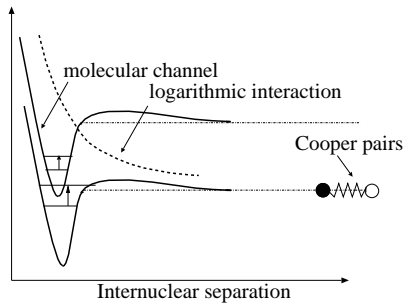}
\caption{Schematic showing the modification of the interatomic
potential by an additional logarithmic contribution coming from the
global vortex structure. Figure is not drawn to scale.}
\label{shaperes}
\end{figure}

Let us now assume that there is a Feshbach resonance in the $s$-wave
interaction between atoms in the up and down states. We argue in this
letter that this can generate a crossover in the FQH regime between
two different types of strongly correlated many-body states.  The
resonant interaction persists even in the gauge transformed composite
particle picture \cite{bhongale}. However,the additional two-body
contributions from Eq.~(\ref{h2}), which have the character of a
repulsive logarithmic potential, can strongly modify the resonance
properties. In a single-channel picture of the Feshbach mechanism,
which is accurate for broad resonances, the additional potential will
shift the bound states in the potential to higher energies, and
therefore also shift the crossover region as shown in
Fig.~\ref{phasedig}. In a two-channel picture of the Feshbach
mechanism (see Fig.~\ref{shaperes}), valid for narrow resonances, the
same shift of the molecular levels in the so-called closed channels
will take place. However, the scattering as a whole could be more
involved as the scattering properties of the open channel can be
affected as well.  Thus the strong correlations induce modifications
to the crossover physics and can be treated systematically within a
Chern-Simons composite particle picture once the details of the
two-body inter atomic potential are known.

We will now consider a specific FQH state to investigate the nature of
the crossover. If we are on the weakly attractive side, then in the
FQH regime the situation is analogous to the one described by the
Haldane-Rezayi (HR) paired wavefunction \cite{haldane}
\begin{eqnarray}
\Psi_{HR}&=&\text{det}(\Xi)\prod_{i<j}(z_i-z_j)^2
\prod_{i<j}(\xi_i-\xi_j)^2 \prod_{i,j}(z_i-\xi_j)^2 \nonumber\\
&&\times \exp\left[-\sum_k |z_k|^2/4-\sum_k
|\xi_k|^2/4\right],\label{HR}
\end{eqnarray}
where $z$ and $\xi$ are scaled in units of harmonic oscillator length
$l_0$ and represent the complex coordinate of the spin up and spin
down components respectively, and $\Xi$ represents pairing between the
up and down particles. Apart from the pairing part, the above
wavefunction for our atomic system is completely justified due to the
short range repulsive piece of the interatomic interaction. The
anti-symmetrization under exchange of identical spin components is
taken care of by the determinant of $\Xi$.

The form of $\Xi$ will turn out to be important in understanding the
nature of the crossover. For this purpose we again work in the
Chern-Simons gauge transformed picture with $\nu=2$ for the HR state
of Eq.~(\ref{HR}). By virtue of this transformation the fermionic
atoms are transformed to free interacting composite particles (bound
state composed of a fermion and 2-vortices) possessing Fermi
statistics. We can therefore write the Hamiltonian for the composite
fermion (CF) system in the standard BCS form
\begin{eqnarray}
H_{CF}=\sum_{\sigma,{\bf k}}(\epsilon_k-\mu)a_{\sigma,{\bf
k}}^{\dagger}a_{\sigma,{\bf
k}}+\hspace{3cm}\nonumber&&
\\\hspace{.5cm}\sum_{{\bf q},{\bf k},{\bf
k}'}U_{{\bf q},{\bf k},{\bf k}'}a_{\uparrow{\bf q}/2+{\bf
k}}^{\dagger}a_{\downarrow{\bf q}/2-{\bf
k}}^{\dagger}a_{\downarrow{\bf q}/2-{\bf k}'}a_{\uparrow{\bf q}/2+{\bf
k}'}.&&
\end{eqnarray}
Here $a_{\sigma,{\bf k}}$ and $a^{\dagger}_{\sigma,{\bf k}}$ are
annihilation and creation operators for composite fermions with
momentum ${\bf k}$ and spin $\sigma$ respectively, $\mu$ is the
chemical potential and $U$ represents the effective attractive inter
composite particle interaction. We diagonalize the above Hamiltonian
via Bogoliubov transformations $\gamma_{{\bf k}\uparrow}=u_{\bf
k}a_{{\bf k}\uparrow}-v_{\bf k}a_{-{\bf k}\downarrow}^{\dagger}$ and
$\gamma_{-{\bf k}\downarrow}^{\dagger}=u_{\bf k}a_{-{\bf
k}\downarrow}^{\dagger}+v_{\bf k}a_{{\bf k}\uparrow}$, $
H_{CF}=\sum_{{\bf k},\sigma}E_k\gamma_{{\bf
k}\sigma}^{\dagger}\gamma_{{\bf k}\sigma} $ with $E_{\bf
k}=\sqrt{(\epsilon_{\bf k}-\mu)^2+\Delta_{\bf k}^2}$, $u_{\bf
k}^2=(1/2)(1+(\epsilon_{\bf k}-\mu)/E_{\bf k})$, and $v_{\bf
k}^2=(1/2)(1-(\epsilon_{\bf k}-\mu)/E_{\bf k})$.  If we now define
$g_{\bf k}=v_{\bf k}/u_{\bf k}$, then the configuration space first
quantized wavefunction for $2N=N_{\uparrow}+N_{\downarrow}$ composite
fermions can be written as $\Psi_{CF}({\bf x_1},{\bf x_2},..{\bf
x_{2N}})=\langle 0|\hat{\psi}({\bf x_{2N}})...\hat{\psi}({\bf
x_{2}})\hat{\psi}({\bf x_{1}})|G\rangle={\mathscr
A}(\phi_{11'}\phi_{22'}...\phi_{NN'})\label{npartwave}$, where
$|G\rangle$ is the variational second quantized BCS wavefunction,
primed and unprimed indexes correspond to different spin components
and $\phi_{jj'}=\sum_{\bf k}g_{\bf k}e^{i{\bf k}\cdot({\bf x}_j-{\bf
x}_{j'})}$, and the anti-symmetrization is separately performed over
up and down spins \cite{schrieffer}. Thus one can identify $\Xi$ in
Eq.~(\ref{HR}) with the product of $\phi$'s over different pairs, and
the anti-symmetrization operator $\mathscr{A}$ with the determinant.

In the BCS theory $E_{\bf k}$ and $\Delta_{\bf k}$ are found self
consistently from the gap and the number equation. Here we will not do
such a calculation, however only focus on the nature of pairing
phases.  In the ${\bf k}\rightarrow 0$ limit, the weak pairing phase
corresponds to $\epsilon_{\bf k}-\mu<0$ where $|u_{\bf k}|\rightarrow
0$ and $|v_{\bf k}|\rightarrow 1$. Thus the leading behavior of
$g_{\bf k}$ goes as $1/u_{\bf k}\propto 1/\Delta_{\bf k}$. However in
the BCS phase $\Delta_{\bf k}$ is significant only in the immediate
vicinity of the Fermi surface, $\phi_{jj'}$ acquires a long range
exponential tail. Thus it is reasonable to assume an exponentially
decaying form $\phi_{jj'}=e^{-|z_j-z_{j'}|/\eta}$ for the pairing
function where $\eta \equiv\hbar v_f/(\pi\Delta_0)$ is the BCS
coherence length and $v_f$ is the Fermi velocity.

\begin{figure}[t]
  \includegraphics[scale=.40]{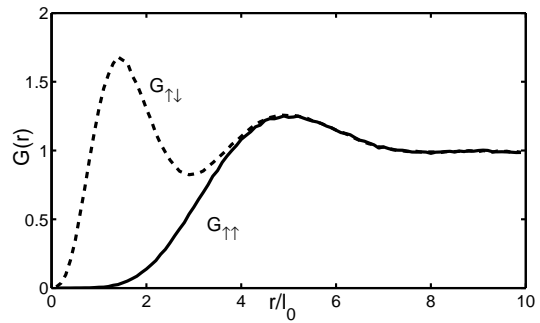}
\caption{Figure shows the two particle correlation functions
$G_{\uparrow\uparrow}({\bf r})$ and $G_{\uparrow\downarrow}({\bf r})$
for $N_{\uparrow}=N_{\downarrow}=100$ for $\eta=l_0$. For large
$r/l_0$, $G_{\uparrow\uparrow}({\bf r})-G_{\uparrow\downarrow}({\bf
r})\rightarrow 0$.}
\label{corr1}
\end{figure}

Now as the strength of interaction is increased by tuning towards
resonance resulting in stronger pairing, the gap $\Delta_0$ increases
exponentially and one may argue that the BCS description is no longer
valid. However as mentioned before we are only concerned about the
form of the pairing wavefunction. Let us therefore consider the
$s$-wave $T$ matrix instead which in the $k\rightarrow 0$ limit is a
smooth function of $a_s$, the s-wave scattering length
\cite{servaas}. We would like to point out here that even though FQH
effect exists in 2D systems, ultra-cold atomic systems under extreme
rotations can be considered to be quasi-2D. Quasi here means that the
confinement in the third dimension is strong compared to the remaining
two. Therefore the scattering can still be considered to be in 3D
justifying the use of the particular $T$ matrix above.  Thus we notice
that even near the Feshbach resonance, the functional form of the $T$
matrix and hence the gap $\Delta_{\bf k}$ will remain unchanged
hinting a smooth crossover. Thus we will parametrize the crossover by
the ratio $\eta/l_0$. This is quite different from the case where the
pairing interaction is $p$-wave where $\Delta_{\bf
k}=\tilde{\Delta}(k_x+ik_y)$. There even if the functional form of the
$T$ matrix remains unchanged, the extra phase associated with the
$\Delta_{\bf k}$ can give a totally different behavior for $g_{\bf k}$
in the strong and weak pairing limits. As argued by Read and Green,
the $p$-wave paired FQH state in the strong and weak pairing limits is
separated by a second order phase transition \cite{read}.

Having obtained the form of the paired FQH state as a function of
$\eta$ we can directly calculate the two particle correlation function
$G({\bf r}_1-{\bf r}_2)=\int..\int d^2{\bf r}_3 ..d^2{\bf r}_N
|\Psi_{HR}|^2$ for different values of $\eta$ by using a metropolis
Monte-Carlo algorithm with $N_{\uparrow}=N_{\downarrow}=100$. In
Fig.~\ref{corr1}, we plot both $G_{\uparrow\uparrow}({\bf r})$ and
$G_{\uparrow\downarrow}({\bf r})$ for $\eta/l_0=1$. We see that
$G_{\uparrow\downarrow}({\bf r})$ shows a peaked behavior for small
$r$ that is absent in $G_{\uparrow\uparrow}({\bf r})$. At the same
time for large $r$, $G_{\uparrow\uparrow}({\bf
r})-G_{\uparrow\downarrow}({\bf r})\rightarrow 0$ implying the
existence of a sum rule special to the HR state valid throughout the
region of our current interest.

\begin{figure}[t]
  \includegraphics[scale=.40]{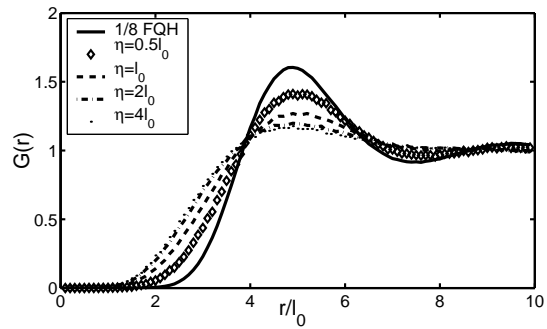}
\caption{Figure shows the comparison between the correlation function
$G_{\uparrow\uparrow}({\bf r})$ in the strong pairing limit and the
$G({\bf r})$ for the $1/8$ FQH wavefunction Eq.~(\ref{laughwave}).}
\label{corr2}
\end{figure}

The crossover behavior is clear from Fig.~\ref{corr2}, which shows
that as $\eta$ becomes small compared to $l_0$,
$G_{\uparrow\uparrow}({\bf r})$ gets modified continuously and tends
towards a limiting form. However the most important point to note is
that the limiting form of $G_{\uparrow\uparrow}({\bf r})$ is exactly
that of the $G({\bf r})$ for the $(1/8)$-FQH state given by the
Laughlin form \cite{laughlin}
\begin{equation}
\Psi_{1/8}=\prod_{i<j}(z_i-z_j)^8
\exp\left[-\sum_k|z_k|^2/2\right].\label{laughwave}
\end{equation}
One way to understand this transition is as follows. In the $\nu=2$ HR
state each composite fermion is associated with two vortices
(quanta). Therefore a molecule formed out of two composite fermions
will consist of four quantas. Moreover, since the molecule has twice
the mass the molecular harmonic oscillator length is $l_0/\sqrt{2}$
and hence eight quanta are required and therefore the fraction is
$1/8$ for the bosonic Laughlin state. Even though our calculations are
for the particular HR state, it is justified to expect that similar
behavior will be obtained for other strongly correlated states that
occur at slightly lower rotational frequencies between the FQH and the
vortex lattice phase.

In conclusion, we have shown that the strong correlations associated
with rapid rotations can cause strong modifications to the crossover,
for example shift the crossover towards the BEC side relative to the
non-rotating case. Using the example of the HR wavefunction we have
shown that the crossover is smooth and the paired FQH state of
fermions smoothly goes over to $1/8$ bosonic FQH state of molecules
when one goes across the Feshbach resonance so that $\eta\ll l_0$.

A detailed calculation of the crossover physics of this region will
require an elaborate treatment with the effect of rotations included
by an effective potential in addition to the actual multichannel
interatomic potential. Within such an effective picture
Nozi\`{e}res-Schmitt-Rink calculations of the crossover region
\cite{nsr} can be carried out. Also these calculations can be extended
to situations with $p$- and $d$- wave pairing schemes in ultra-cold
Fermi gases. These scenarios while having close resemblance with, for
example, the 5/2 FQH effect, will be extremely useful and will be
dealt with in a future publication.

At the same time paired FQH states such as $5/2$ are known to possess
exotic non-abelian quasi particles excitation. While the existence of
non-abelian statistics is the basis for topological scheme of
implementing quantum logic in a quantum computer, the $5/2$ state is
known to be computationally non-universal. However, there have been
proposals \cite{bravyi} in which this symptom can be remedied by
dynamically tuning-in additional non-topological interactions. Dynamic
control while hard in the solid state configurations of the FQH
effect, controlled transitions between different FQH states like the
one discussed here may be extremely useful for implementing such
topological schemes.

We thank Anil Shaji and Maikel Goosen for useful discussions. SB
acknowledges financial support from the ONR, Contract
No. N00014-03-1-0508. SK acknowledges financial support from the NWO.

\bibliography{cross}

\begin{thebibliography}{20}
\expandafter\ifx\csname natexlab\endcsname\relax\def\natexlab#1{#1}\fi
\expandafter\ifx\csname bibnamefont\endcsname\relax
  \def\bibnamefont#1{#1}\fi
\expandafter\ifx\csname bibfnamefont\endcsname\relax
  \def\bibfnamefont#1{#1}\fi
\expandafter\ifx\csname citenamefont\endcsname\relax
  \def\citenamefont#1{#1}\fi
\expandafter\ifx\csname url\endcsname\relax
  \def\url#1{\texttt{#1}}\fi
\expandafter\ifx\csname urlprefix\endcsname\relax\def\urlprefix{URL }\fi
\providecommand{\bibinfo}[2]{#2}
\providecommand{\eprint}[2][]{\url{#2}}

\bibitem[{\citenamefont{Feshbach}(1958)}]{fesh1}
\bibinfo{author}{\bibfnamefont{H.}~\bibnamefont{Feshbach}},
  \bibinfo{journal}{Ann. Phys.} \textbf{\bibinfo{volume}{5}},
  \bibinfo{pages}{357} (\bibinfo{year}{1958}).

\bibitem[{\citenamefont{Feshbach}(1962)}]{fesh2}
\bibinfo{author}{\bibfnamefont{H.}~\bibnamefont{Feshbach}},
  \bibinfo{journal}{Ann. Phys.} \textbf{\bibinfo{volume}{19}},
  \bibinfo{pages}{287} (\bibinfo{year}{1962}).

\bibitem[{\citenamefont{Greiner et~al.}(2003)\citenamefont{Greiner, Regal, and
  Jin}}]{fermicond1}
\bibinfo{author}{\bibfnamefont{M.}~\bibnamefont{Greiner}},
  \bibinfo{author}{\bibfnamefont{C.}~\bibnamefont{Regal}}, \bibnamefont{and}
  \bibinfo{author}{\bibfnamefont{D.}~\bibnamefont{Jin}},
  \bibinfo{journal}{Nature} \textbf{\bibinfo{volume}{426}},
  \bibinfo{pages}{537} (\bibinfo{year}{2003}).

\bibitem[{\citenamefont{Jochim et~al.}(2003)\citenamefont{Jochim, Bartenstein,
  Altmeyer, Hendl, Riedl, Chin, Denschlag, and Grimm}}]{fermicond2}
\bibinfo{author}{\bibfnamefont{S.}~\bibnamefont{Jochim}},
  \bibinfo{author}{\bibfnamefont{M.}~\bibnamefont{Bartenstein}},
  \bibinfo{author}{\bibfnamefont{A.}~\bibnamefont{Altmeyer}},
  \bibinfo{author}{\bibfnamefont{G.}~\bibnamefont{Hendl}},
  \bibinfo{author}{\bibfnamefont{S.}~\bibnamefont{Riedl}},
  \bibinfo{author}{\bibfnamefont{C.}~\bibnamefont{Chin}},
  \bibinfo{author}{\bibfnamefont{J.~H.} \bibnamefont{Denschlag}},
  \bibnamefont{and} \bibinfo{author}{\bibfnamefont{R.}~\bibnamefont{Grimm}},
  \bibinfo{journal}{Science} \textbf{\bibinfo{volume}{302}},
  \bibinfo{pages}{2101} (\bibinfo{year}{2003}).

\bibitem[{\citenamefont{Zwierlein et~al.}(2003)\citenamefont{Zwierlein, Stan,
  Schunch, Raupach, Gupta, Hadzibabic, and Ketterle}}]{fermicond3}
\bibinfo{author}{\bibfnamefont{M.~M.} \bibnamefont{Zwierlein}},
  \bibinfo{author}{\bibfnamefont{C.~A.} \bibnamefont{Stan}},
  \bibinfo{author}{\bibfnamefont{C.~H.} \bibnamefont{Schunch}},
  \bibinfo{author}{\bibfnamefont{S.}~\bibnamefont{Raupach}},
  \bibinfo{author}{\bibfnamefont{S.}~\bibnamefont{Gupta}},
  \bibinfo{author}{\bibfnamefont{Z.}~\bibnamefont{Hadzibabic}},
  \bibnamefont{and} \bibinfo{author}{\bibfnamefont{W.}~\bibnamefont{Ketterle}},
  \bibinfo{journal}{Phys. Rev. Lett.} \textbf{\bibinfo{volume}{91}},
  \bibinfo{pages}{250401} (\bibinfo{year}{2003}).

\bibitem[{\citenamefont{Eagles}(1969)}]{eagles}
\bibinfo{author}{\bibfnamefont{D.~M.} \bibnamefont{Eagles}},
  \bibinfo{journal}{Phys. Rev.} \textbf{\bibinfo{volume}{186}},
  \bibinfo{pages}{456} (\bibinfo{year}{1969}).

\bibitem[{\citenamefont{Leggett}(1980)}]{leggett}
\bibinfo{author}{\bibfnamefont{A.~J.} \bibnamefont{Leggett}},
  \bibinfo{journal}{J. Phys. Colloq.} \textbf{\bibinfo{volume}{41}},
  \bibinfo{pages}{7} (\bibinfo{year}{1980}).

\bibitem[{\citenamefont{Zwierlein et~al.}(2005)\citenamefont{Zwierlein,
  Abo-Shaeer, Schirotzek, Schunck, and Ketterle}}]{crossvert}
\bibinfo{author}{\bibfnamefont{M.~W.} \bibnamefont{Zwierlein}},
  \bibinfo{author}{\bibfnamefont{J.~R.} \bibnamefont{Abo-Shaeer}},
  \bibinfo{author}{\bibfnamefont{A.}~\bibnamefont{Schirotzek}},
  \bibinfo{author}{\bibfnamefont{C.~H.} \bibnamefont{Schunck}},
  \bibnamefont{and} \bibinfo{author}{\bibfnamefont{W.}~\bibnamefont{Ketterle}},
  \bibinfo{journal}{Nature} \textbf{\bibinfo{volume}{435}},
  \bibinfo{pages}{1047} (\bibinfo{year}{2005}).

\bibitem[{\citenamefont{Wilkins and Gunn}(2000)}]{wilkins}
\bibinfo{author}{\bibfnamefont{K.}~\bibnamefont{Wilkins}} \bibnamefont{and}
  \bibinfo{author}{\bibfnamefont{J.~M.~F.} \bibnamefont{Gunn}},
  \bibinfo{journal}{Phys. Rev. Lett} \textbf{\bibinfo{volume}{84}},
  \bibinfo{pages}{6} (\bibinfo{year}{2000}).

\bibitem[{\citenamefont{Paredes et~al.}(2001)\citenamefont{Paredes, Fedichev,
  Cirac, and Zoller}}]{zoller}
\bibinfo{author}{\bibfnamefont{B.}~\bibnamefont{Paredes}},
  \bibinfo{author}{\bibfnamefont{P.}~\bibnamefont{Fedichev}},
  \bibinfo{author}{\bibfnamefont{J.~I.} \bibnamefont{Cirac}}, \bibnamefont{and}
  \bibinfo{author}{\bibfnamefont{P.}~\bibnamefont{Zoller}},
  \bibinfo{journal}{Phys. Rev. Lett.} \textbf{\bibinfo{volume}{87}},
  \bibinfo{pages}{010402} (\bibinfo{year}{2001}).

\bibitem[{\citenamefont{Zhang}(1992)}]{zhang}
\bibinfo{author}{\bibfnamefont{S.~C.} \bibnamefont{Zhang}},
  \bibinfo{journal}{Int. J. Mod. Phys. B} \textbf{\bibinfo{volume}{6}},
  \bibinfo{pages}{25} (\bibinfo{year}{1992}).

\bibitem[{\citenamefont{Jacak et~al.}(2003)\citenamefont{Jacak, Sitko,
  Wieczorek, and W\'{o}js}}]{lucjan}
\bibinfo{author}{\bibfnamefont{L.}~\bibnamefont{Jacak}},
  \bibinfo{author}{\bibfnamefont{P.}~\bibnamefont{Sitko}},
  \bibinfo{author}{\bibfnamefont{K.}~\bibnamefont{Wieczorek}},
  \bibnamefont{and} \bibinfo{author}{\bibfnamefont{A.}~\bibnamefont{W\'{o}js}},
  \emph{\bibinfo{title}{Quantum Hall Systems}} (\bibinfo{publisher}{Oxford
  University Press}, \bibinfo{year}{2003}), chap.~\bibinfo{chapter}{7}.

\bibitem[{\citenamefont{Bhongale et~al.}(2004)\citenamefont{Bhongale, Milstein,
  and Holland}}]{bhongale}
\bibinfo{author}{\bibfnamefont{S.~B.} \bibnamefont{Bhongale}},
  \bibinfo{author}{\bibfnamefont{J.~N.} \bibnamefont{Milstein}},
  \bibnamefont{and} \bibinfo{author}{\bibfnamefont{M.~J.}
  \bibnamefont{Holland}}, \bibinfo{journal}{Phys. Rev. A}
  \textbf{\bibinfo{volume}{69}}, \bibinfo{pages}{053603}
  (\bibinfo{year}{2004}).

\bibitem[{\citenamefont{Haldane and Rezayi}(1988)}]{haldane}
\bibinfo{author}{\bibfnamefont{F.~D.~M.} \bibnamefont{Haldane}}
  \bibnamefont{and} \bibinfo{author}{\bibfnamefont{E.~H.}
  \bibnamefont{Rezayi}}, \bibinfo{journal}{Phys. Rev. Lett.}
  \textbf{\bibinfo{volume}{60}}, \bibinfo{pages}{956} (\bibinfo{year}{1988}).

\bibitem[{\citenamefont{Schrieffer}(1964)}]{schrieffer}
\bibinfo{author}{\bibfnamefont{J.~R.} \bibnamefont{Schrieffer}},
  \emph{\bibinfo{title}{Theory of Superconductivity}}
  (\bibinfo{publisher}{W.~A. Benjamin Inc., Publishers}, \bibinfo{address}{New
  York}, \bibinfo{year}{1964}).

\bibitem[{\citenamefont{Kokkelmans et~al.}(2002)\citenamefont{Kokkelmans,
  Milstein, Chiofalo, Walser, and Holland}}]{servaas}
\bibinfo{author}{\bibfnamefont{S.~J.~J.~M.~F.} \bibnamefont{Kokkelmans}},
  \bibinfo{author}{\bibfnamefont{J.~N.} \bibnamefont{Milstein}},
  \bibinfo{author}{\bibfnamefont{M.~L.} \bibnamefont{Chiofalo}},
  \bibinfo{author}{\bibfnamefont{R.}~\bibnamefont{Walser}}, \bibnamefont{and}
  \bibinfo{author}{\bibfnamefont{M.~J.} \bibnamefont{Holland}},
  \bibinfo{journal}{Phys. Rev. A} \textbf{\bibinfo{volume}{65}},
  \bibinfo{pages}{053617} (\bibinfo{year}{2002}).

\bibitem[{\citenamefont{Read and Green}(2000)}]{read}
\bibinfo{author}{\bibfnamefont{N.}~\bibnamefont{Read}} \bibnamefont{and}
  \bibinfo{author}{\bibfnamefont{D.}~\bibnamefont{Green}},
  \bibinfo{journal}{Phys. Rev. B} \textbf{\bibinfo{volume}{61}},
  \bibinfo{pages}{10267} (\bibinfo{year}{2000}).

\bibitem[{\citenamefont{Laughlin}(1983)}]{laughlin}
\bibinfo{author}{\bibfnamefont{R.~B.} \bibnamefont{Laughlin}},
  \bibinfo{journal}{Phys. Rev. Lett.} \textbf{\bibinfo{volume}{50}},
  \bibinfo{pages}{1395} (\bibinfo{year}{1983}).

\bibitem[{\citenamefont{Nozi\`{e}res and Schmitt-Rink}(1985)}]{nsr}
\bibinfo{author}{\bibfnamefont{P.}~\bibnamefont{Nozi\`{e}res}}
  \bibnamefont{and}
  \bibinfo{author}{\bibfnamefont{S.}~\bibnamefont{Schmitt-Rink}},
  \bibinfo{journal}{J. Low. Temp. Phys.} \textbf{\bibinfo{volume}{59}},
  \bibinfo{pages}{195} (\bibinfo{year}{1985}).

\bibitem[{\citenamefont{Bravyi}(2006)}]{bravyi}
\bibinfo{author}{\bibfnamefont{S.}~\bibnamefont{Bravyi}},
  \bibinfo{journal}{Phys. Rev. A} \textbf{\bibinfo{volume}{73}},
  \bibinfo{pages}{042313} (\bibinfo{year}{2006}).

\end{thebibliography}

\end{document}